\documentclass[twocolumn,showpacs,nofootinbib,tightenlines,nobibnotes, aps, amssymb,amsmath,prl]{revtex4}
\usepackage{graphicx}
\usepackage{epsfig}

\newcommand{\be}{\begin{eqnarray}}
\newcommand{\ee}{\end{eqnarray}}

\begin{document}
\title{Universal relaxational dynamics of gapped one dimensional models in the quantum
sine-Gordon universality class}
\author{Kedar Damle$^1$ and Subir Sachdev$^2$}
\affiliation{$^1$Department of Theoretical Physics,
Tata Institute of Fundamental Research, Mumbai 400005, India\\
$^2$Department of Physics, Harvard University, Cambridge, MA
02138}

\date{\today}

\begin{abstract}
A semiclassical approach to the low-temperature real time dynamics
of generic one-dimensional, gapped models in the sine-Gordon model
universality class is developed. Asymptotically exact universal
results for correlation functions are obtained in the temperature
regime $T \ll \Delta$, where $\Delta$ is the energy gap.
\end{abstract}

\vskip2pc

\maketitle

Real time, non-zero temperature ($T$), correlation functions of
quantum many body systems are directly related to many
experimentally measurable quantities.   For strongly interacting
systems, there are few quantitative results on the relaxation and
transport processes that are believed to occur at long times at
any $T\neq 0$. Standard perturbative methods, as well as many
numerical approaches, work best in imaginary time, but the
analytic continuation of such imaginary time results to real time
is most dangerous, and often fails, in the low frequency limit.
Among the few exact results available are those describing
relaxational behaviour of an order parameter at conformally
invariant critical points (obtained by exploiting the conformal
invariance of the system), or by semiclassical methods deep in an
ordered state~\cite{SachdevYoung}, both in dimension $d=1$. In an
important recent development, Altshuler and Tsvelik
\cite{AltshulerTsvelik} (AT) obtained real time order parameter
correlations for the integrable $d=1$ quantum Ising model by the
form-factor expansion, and their results were in precise agreement
with the semiclassical predictions \cite{SachdevYoung}.

AT also argued that their results could be extended to other
gapped, integrable systems, and in particular, to the sine-Gordon
model; some results for another massive integrable model (the O(3)
non-linear sigma model) had also been obtained earlier by
Fujimoto~\cite{Fujimoto} and Konik~\cite{Konik}. In this paper, we
obtain dynamic non-zero $T$ correlators of gapped one dimensional
models in the universality class of the $d=1$ quantum sine-Gordon
field theory using a semiclassical method~\cite{SachdevYoung}: our
results are expected to apply generic, non-integrable models \cite{DamleSachdev}.
Interestingly, as we describe in detail below, our semiclassical
results are {\em qualitatively different} from those of AT. One
likely origin of this difference is that the long time transport
and spectral~\cite{Huse} properties of integrable models,
with infinitely many conserved
charges~\cite{EsslerKonik}, are genuinely distinct from those of
more generic models; the latter have only a small number of
conserved charges and are described by the semiclassical theory.
The generic quantum Ising model has only a single conserved charge
(the energy density) and was possibly too simple a model to expose
the above distinction. At the end of this paper, we will also
mention other possible sources of the difference between our
sine-Gordon results and those of AT and others
\cite{Fujimoto,Konik}.

The imaginary time ($\tau$) action of the sine-Gordon model is
\begin{displaymath}
{\cal A} = \frac{c}{16\pi} \int_0^{1/T} d\tau dx \left[ (\partial_x \Phi
)^2 + \frac{1}{c^2}(\partial_{\tau} \Phi)^2 - g^2\cos(\gamma \Phi)  \right]
\end{displaymath}
where $\Phi$ is the sine-Gordon field, and $c$ has dimensions of
velocity. It is assumed that the full integrability of the model
is broken either by additional terms not shown above, or by an
ultraviolet cutoff that does not preserve integrability. Such
perturbations are always present in any experimentally relevant
realization (see Ref.~\onlinecite{EsslerKonik} for examples of experimental
applications), but will be left implicit in our discussion below; we
will thus freely refer to {\em the} sine-Gordon model when we mean
generic models in this universality class. In the integrable
sine-Gordon theory, $g$ is relevant for $\gamma < 1$, giving rise
to a gapped phase with particle-like excitations. For $1/\sqrt{2}
< \gamma < 1$, the interactions between these particles are purely
repulsive and there are no bound states. It is expected that a
similar `purely repulsive' regime exists without full
integrability.

Our focus here is on the space and time dependent correlator of
$e^{i\eta\Phi}$ defined in the usual way for arbitrary $\eta$:
$C_{\Phi}(x,t) \equiv \left\langle e^{ i\eta\Phi (x,t)} e^{- i\eta \Phi (0,0)} \right\rangle $, where $\Phi(x,t) \equiv e^{i\hat{\cal H}t} \Phi(x)
e^{-i\hat{\cal H}t}$, $\hat{\cal H}$ is the Hamiltonian of the system, and the angular
brackets indicate average over the equilibrium density matrix of
the system. Below, we shall obtain the  space-time dependent
$C(x,t)$ for temperatures $T \ll \Delta$ ($\Delta$ is the gap) in
the purely repulsive regime of the gapped phase mentioned above.
Our answer will be a universal scaling function of $x$, $t$,
$\Delta$ and $c$, a velocity that enters the dispersion relation
of the solitonic excitations.

There are three key observations that allow our exact computation
for $T\ll \Delta$. The first is that as there is an excitation
gap, and the lowest lying excitations above the gap are `charged'
particles (with charges $\pm 1$) that represent solitons and
anti-solitons in the sine-Gordon field. For $T \ll \Delta$, the
particle dispersion $\epsilon(p)$ may be set equal to its
universal low momentum form $\epsilon(p) = \Delta +
p^2c^2/2\Delta$, and they may be treated semiclassicaly using
Maxwell-Boltzmann statistics with a velocity probability density
$P(v) \sim e^{-\Delta v^2/2 c^2 T}$. Indeed, their density
\begin{equation}
\rho = 2 e^{-\Delta/T} \int \frac{dp}{2\pi} e^{-p^2c^2/(2\Delta
T)}, \label{densityvalue}
\end{equation}
while their r.m.s. velocity $v$ is $v_T = c (T/\Delta)^{1/2}$. The
mean inter-particle spacing $\sim e^{+\Delta/T}$ is thus much
larger than the thermal de-Broglie wavelength $\sim c/(\Delta
T)^{-1/2}$, and the particles behave
semiclassically~\cite{SachdevYoung}.

\begin{figure}
\epsfxsize=2.5in \centerline{\epsffile{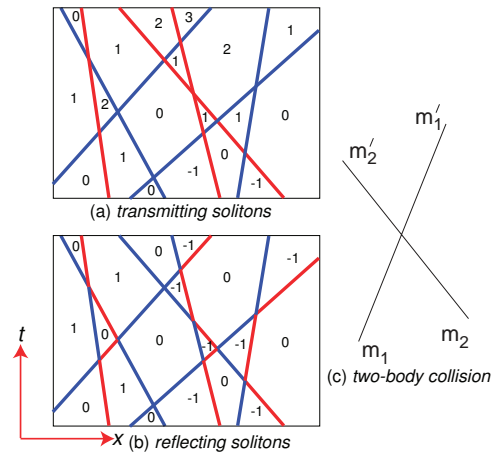}}
\caption{A typical set of particle trajectories contributing to
$C_\Phi (x,t)$. Each trajectory represents paths moving both
forward and backward in time.
The charge moving on a trajectory is indicated by different colors (color
online): Red corresponds to
solitons ($m=+1$) and blue to anti-solitons ($m=-1$). The numbers in the
domains are the local values of the integer $n$ in Eq.~(\ref{nn})}
\label{plotF}
\end{figure}

The second observation is that collisions between these particles are
described by their two-particle $S$-matrix, and only a
simple universal low-velocity limit of this $S$-matrix is needed in the low
$T$ limit since the particle `rapidity' $\sim v_T / c \ll 1$.
In this limit, in the purely repulsive regime of the gapped phase,
the $S$-matrix
for the process in
Fig~\ref{plotF}c is generically~\cite{EsslerKonik}
\begin{equation}
 {\cal S}^{m_1 m_2}_{m^{\prime}_1,
m^{\prime}_2} = (-1) \delta_{m_1 m^{\prime}_2} \delta_{m_2 m^{\prime}_1}.
\label{smatrix}
\end{equation}
In other words, the excitations behave like impenetrable
particles which preserve their charge in a collision.
Energy and momentum conservation in $d=1$ require that these
particles simply exchange momenta across a collision
(Fig~\ref{plotF}c). The $(-1)$ factor in Eq.~(\ref{smatrix}) can be
interpreted as the phase-shift of repulsive scattering between
slowly moving bosons in $d=1$. Indeed, the simple generic form of
Eq.~(\ref{smatrix}) is due to the slow motion of the
particles \cite{DamleSachdev}, and is
not a special feature of relativistic continuum theory. It is also
worth noting that at special {\em free fermion points} in the
gapped phase (when a mapping to non-interacting fermions exists),
the repulsive interaction between the solitonic particles goes to
zero and the $S$-matrix is then perfectly transmitting {\em i.e.\/}  it
equals $(-1) \delta_{m_1
m^{\prime}_1} \delta_{m_2 m^{\prime}_2}$ (for
generic, non-integrable models, accessing this free fermion
behavior requires tuning an infinite number of couplings, and so
this point is not physically relevant). The dynamics in this
special case is expected to be very different from the generic
behavior, and will be discussed below as a prelude to our results
in the generic case.

Finally, we note that in the semiclassical limit at low
temperatures, for most of the time, the sine-Gordon field $\Phi$
takes one of the values
\begin{equation}
\Phi = \frac{2 \pi n}{\gamma} \label{nn}
\end{equation}
where $n$ is an integer which increases (decreases) upon crossing
in the positive $x$ direction a soliton (anti-soliton) trajectory.
A sample set of soliton trajectories in spacetime are shown in
Fig~\ref{plotF}, along with the associated values of $n$.
Notice the difference in the $n$ values between the transmitting
soliton free-fermion case, and the generic reflecting soliton case
described by Eq.~(\ref{smatrix}).

We now present our final results for $C_\Phi (x,t)$, deferring
their derivation till later. For the transmitting soliton case in
Fig~\ref{plotF}a, an elementary computation shows that
\begin{equation}
C_\Phi (x,t) = {\cal A} \exp \left( - 2 \rho \sin^2 (\pi \eta/\gamma)
\left\langle |x - v t| \right\rangle \right), \label{e5}
\end{equation}
where the angular brackets represent an average of $v$ over
$P(v)$ and ${\cal A}$ is an $\eta$ and $\gamma$ dependent
amplitude~\cite{AltshulerTsvelik}
related to the vacuum expectation value of $e^{i\eta \Phi}$. Remarkably, Eq.~(\ref{e5}) is precisely the result obtained by
AT for the integrable sine-Gordon model for generic values of
$\gamma$ even away from the free-fermion point. So it appears that
the dynamics of the integrable model is rather like that of
non-interacting particles.

For generic reflecting soliton case in Fig~\ref{plotF}b, with the $S$-matrix
in Eq.~(\ref{smatrix}) (which we
claim is always the experimentally relevant case), we obtained a
very different result. We found
\begin{eqnarray}
&& C_\Phi(x,t) = {\cal A} e^{-(\tilde{q}_r+\tilde{q}_l)}
\Biggl[U_0(2i\tilde{q}_r\Theta,2i\sqrt{\tilde{q}_r\tilde{q}_l}) +
\nonumber \\
&& U_0(2i\tilde{q}_l\Theta,2i\sqrt{\tilde{q}_r\tilde{q}_l})
-iU_1(2i\tilde{q}_r\Theta,2i\sqrt{\tilde{q}_r\tilde{q}_l})
\nonumber \\ &&
                  -iU_1(2i\tilde{q}_l\Theta,2i\sqrt{\tilde{q}_r\tilde{q}_l}) -I_0(2\sqrt{\tilde{q}_r\tilde{q}_l})
                  \Biggr], \label{final}
\end{eqnarray}
where $\Theta \equiv \cos(2 \pi \eta/\gamma)$, $\tilde{q}_r = \rho
\int_{-\infty}^{x/t} dv P(v) (x-vt)$, $\tilde{q}_l = \rho
\int_{x/t}^{\infty} dv P(v) (x-vt)$, $I_0$ is the modified Bessel
function, and $U_{0,1}$ are the {\em Lommel} functions of two
variables\cite{WatsonBook}. (Note that all $x$ and and $t$ dependence
is through $\tilde{q}_r$ and $\tilde{q}_l$ which can both
be written as functions of the scaling variables $\bar{x} = \rho x \equiv
x/\xi_x$
and $\bar{t} = c\rho\sqrt{\frac{T}{2\pi\Delta}} t \equiv t/\xi_t$.)

This remarkable expression is {\em qualitatively} different in its
asymptotic behavior for large $t$ and $x$ from the corresponding
result at the free fermion point. One way to see this is to obtain
the correlator of the soliton charge density $\rho = (\gamma/2
\pi) \partial_x \Phi$,  $C_{m}(x,t) \equiv \left \langle \rho(x,t)
\rho(0,0) \right \rangle$, by expanding  Eq.~(\ref{final}) to order
$\eta^2$ and taking two $x$ derivatives. The result is {\em
identical} in form to the {\em diffusive} correlation function
obtained for the conserved (spin) density in
Ref~\onlinecite{DamleSachdev} using the analysis
of Refs~\onlinecite{Jepsen,LebowitzPercus}---here, the
diffusion constant has value $D_m=
\frac{c^2}{2\Delta}e^{\Delta/T}$. A more direct indication of
diffusive behavior follows from an asymptotic analysis of our
result. For instance, for large $t$ and $x \sim \sqrt{t}$, using
$\tilde{q}_{r/l} \simeq t/\xi_t \pm x/2\xi_x$ and standard results for
the asymptotics of Bessel and Lommel functions\cite{WatsonBook},
we find: \be C_\Phi(x,t) & \simeq & {\cal A} \left
(\frac{1+\Theta}{\rho(1-\Theta)} \right )\frac{e^{-x^2/(4D_\Phi
t)}}{(4\pi D_\Phi t)^{1/2}} \label{asymp} \ee where the diffusion
constant $D_{\Phi} = c\sqrt{T}/\rho\sqrt{2\pi \Delta} =
\frac{c^2}{2\Delta}e^{\Delta/T}$ is identical to the diffusion
constant $D_m$ that describes the diffusive behaviour of the
soliton charge correlator. This may be interpreted by noting that
the field values $\Phi(x,t)$ and $\Phi(0,0)$ are perfectly
correlated if the space-time points $(0,0)$ and $(x,t)$ lie in the
{\em same domain}, and this happens when the domain walls bounding
this domain diffuse from the neighbourhood of the origin to the
space-time point $(x,t)$; the equality of the two diffusion
constants then suggests that such events dominate in the average
over all space-time histories. In this context, it is however
important to bear the following subtlety in mind: The asymptotic
expression Eq.~(\ref{asymp}) is only valid as long as $|\Theta | < 1
- x \xi_t/(2t\xi_x)$, and as such cannot be used directly
to expand in $\eta$ about $\eta = 0$ (which of course corresponds
to $\Theta = 1$). Put another way, the long time limit does not
commute with the $\Theta \rightarrow 1$ limit.

We now describe the derivation of Eqs. (\ref{e5}) and
(\ref{final}). We represent $C(x,t)$ as a `double time' path
integral \cite{SachdevYoung}, with the $e^{-i\hat{\cal H}t}$
factor generating {\em trajectories} that move forward in time for
each particle, and the $e^{i \hat{\cal H} t}$ producing
trajectories that move backward in time. In the classical limit,
stationary phase is achieved when the trajectories are
time-reversed pairs of classical paths (Fig~\ref{plotF}). Each
trajectory has a charge label which obeys Eq.~(\ref{smatrix}) at each
collision; however as each collision contributes both to the
forward and backward trajectories, the net numerical factor is
simply $+1$. All of this implies~\cite{SachdevYoung} that the
straight lines ({\em rays}) $y_\mu (t) = y_\mu + v_\mu t$ ($\mu=
1,2,\ldots M$ )in Fig~\ref{plotF} are independently distributed
with a uniform distribution in space for the intercepts $y_\mu$,
and with inverse slope determined by the velocities $v_\mu$ which
are independently distributed according to the probability density
$P(v)$. The charge, $m$, is assigned randomly to each {\em
trajectory} at some initial time $t=0$ with probability
$\rho_{m}/\rho = 1/2$, but then evolves in time as discussed above
(Fig~\ref{plotF}). We have assumed a large system size $L$, and
will eventually take the limit $L \rightarrow \infty$, $M
\rightarrow \infty$ with total density $\rho = M/L$ fixed and
given by Eq.~(\ref{densityvalue}). Note that the charge travelling
on a ray changes in time when the $S$-matrix is reflecting, and
charges on rays are mutually uncorrelated  {\em only at }
$t=0$---it is the charges $m_k$ ($k= 1,2,\ldots M$ labelled
starting from the left) travelling on the complicated zig-zag {\em
trajectories} $x_k(t)$ that remain uncorrelated and constant in time.
Conversely, when the $S$-matrix is perfectly
transmitting at special free-fermion points in the phase diagram,
charges on rays remain uncorrelated and constant in time
(since the {\em rays} and the {\em trajectories} coincide in this
case). In both cases, the field $\Phi (x,t)$ is
given in terms of soliton trajectories by
\begin{equation}
\Phi (x,t) = \frac{2 \pi}{\gamma} \sum_{k=1}  m_k \theta(x - x_k
(t)). \label{e1}
\end{equation}

To proceed further in the generic reflecting case,
we note that the spatial sequence of {\em domains} encountered
at any given time as we move from $x = -\infty$ to $x= +\infty$ is
invariant under the time evolution. Therefore, it is valid to label
the sequence of domains from the left by an index $l$ whereby the
field $\Phi$ takes value $\Phi_l$ independent of $t$ for all $x$
in the $l^{th}$ domain. In doing so, we resolve the singularity
associated with each collision by stipulating that a zero length
domain of the appropriate field value continues to exist {\em at}
the instant of the collision. Now, let space-time points $(x,t)$
and $(0,0)$ lie in the $l_2^{th}$ and $l_1^{th}$ domain
respectively---such an assignment is well-defined precisely
because the sequence of domains is time-invariant. The difference
$\Phi(x,t) - \Phi(0,0)$ is thus equal to $\Phi_{l_2} -
\Phi_{l_1}$. To obtain the latter, it is extremely useful to work
{\em at} $t=0$ and write
\begin{equation}
\Phi_{l_2} - \Phi_{l_1} = \frac{2\pi}{\gamma}\sum_{i=1}^{s}m_i \,
, \label{e6}
\end{equation}
where $m_{1 \dots s}$ are the charges of the $s=|l_2-l_1|$
domain-walls (solitons or antisolitions) encountered when going
from domain $l_2$ to domain $l_1$ {\em at} $t=0$. The advantage
of this procedure is clear: The $m_{1 \dots |l_2-1_1|}$ are uncorrelated
independently random variables that take values $\pm 1$ with
equal probability, and as such, may be averaged over independently
of one another for fixed $s$.

The number $s$ is of course a complicated function of $x$, $t$ and
all the initial positions $y_\mu$ and velocities $v_\mu$ of all
the rays. However, since the $t=0$ soliton charge on a ray is not
correlated with the $y_\mu$ and $v_\mu$, we may perform the
average over these $s$ soliton charges first for a given
realization of the $y_\mu$ and $v_\mu$, and then average the
resulting expression over all the $y_\mu$ and $v_\mu$. The result
of this charge averaging is $\left \langle e^{ i\eta\Phi (x,t)} e^{- i\eta \Phi (0,0)} \right\rangle_{{\mathrm charge}} =  {\cal A} \Theta^s$.

Next, we note that $s$ may be obtained by moving
from the spacetime point $(0,0)$ to point $(x,t)$ along
the straight line joining the two, and keeping track of the numbers
$k_r$ and $k_l$ of rays that cross this straight line from the right
and left respectively. After every such right intersection, we move
one domain to the right, while every left intersection results in
moving back one domain to the left---thus, $s$ is precisely
equal to $|k_r-k_l|$. Since initial positions and velocities of
the rays are independent random variables, the probability of
having $k_r$ right intersections and $k_l$ left intersections
may be calculated straightforwardly by elementary means in terms
of $q_r$, the probability that the straight line connecting point $(0,0)$
to point $(x,t)$ has a right intersection with a {\em given} ray,
and $q_l$, the corresponding probability for left
intersection.
This, in conjunction with the
result for charge averaging above, then allows us to write the following
combinatorial expression
\begin{eqnarray} && C_\Phi (x,t) = {\cal A} \; \; \lim_{M \rightarrow \infty}
\nonumber \\ &&~\sum_{k_r \, , \,
k_l}\frac{M!q_r^{k_r}q_l^{k_l}(1-q_r-q_l)^{M-k_r-k_l}}{k_r! k_l!
(M-k_r-k_l)!}\Theta^{|k_r - k_l|} \, , \label{e8}
\end{eqnarray}
where the primed sum is over all $k_r$ and $k_l$ ranging from $0$
to $M$ with the constraint that $k_r+k_l \leq M$, $q_{l,r} =
\tilde{q}_{l,r}/M$, and we have assumed without loss of generality
that $x$ and $t$ are positive.

Before we proceed further, we note that the perfectly transmitting
case is considerably simpler: Since the trajectories coincide with
rays, and charges on rays are invariant in time and uncorrelated,
$\Phi(x,t) -\Phi(0,0)$ can be obtained by moving along the dotted
line connecting $(0,0)$ and $(x,t)$ and counting the {\em total
number} $k_r+k_l$ of rays intersecting this line (this is the
number of {\em uncorrelated} random charges we cross in going from
$(0,0)$ to $(x,t)$) and multiplying by $2\pi/\gamma$. The rest of
the argument above is unchanged and we obtain a simpler
combinatorial expression with the $|k_r-k_l|$ in the exponent of
the cosine in Eq.~(\ref{e8}) replaced by $k_r+k_l$. Taking the
thermodynamic limit using the trinomial formula gives the result
Eq.~(\ref{e5}).

Returning to the reflecting case, further progress relies
crucially on a device suggested to us by D.~Dhar. The artifice
involved consists of viewing the sum of a subset of terms with
fixed $k_r -k_l = n$ as the coefficient of $z^n$  in the Laurent
expansion of some function of a complex variable $z$. For $n \geq
0$, this is achieved by identifying the corresponding partial sum
with the coefficient of $z^n$ in the trinomial expansion of
$f_M(\Theta z) \equiv (1-q_r-q_l+\Theta zq_r+q_l/\Theta z)^M$.
Similarly, for $n < 0$, the identification is with the coefficient
of $z^n$ in the trinomial expansion of $f_M(z/\Theta )$. The
advantage of this approach is now clear: Each Laurent coefficient
can be written as a contour integral around a contour enclosing
the origin, and the two sums over $n$ for either sign of $n$ are
simply geometric series which can be evaluated in closed form.
Moreover, the thermodynamic limit can be taken explicitly for both
sums upon choosing in each case a contour along which the
geometric series converges. A convenient rescaling of integration
variables then gives:
\begin{eqnarray}
 && C_\Phi(x,t) = {\cal A} \frac{e^{-(\tilde{q}_r+\tilde{q}_l)}}{2\pi i}\left (
\oint_{\mathcal{C}_a} \frac{dw}{w- \Theta
(\frac{\tilde{q}_r}{\tilde{q}_l})^{\frac{1}{2}}}
e^{\sqrt{\tilde{q}_r \tilde{q}_l}(w+w^{-1})} \right. \nonumber \\
&&~~~~~~~~\left. + \oint_{\mathcal{C}_b}
\frac{dw}{w-\frac{1}{\Theta}(\frac{\tilde{q}_r}{\tilde{q}_l})^{\frac{1}{2}}}
e^{\sqrt{\tilde{q}_r \tilde{q}_l}(w+w^{-1})} \right).
\label{contour}
 \end{eqnarray}
Here $\mathcal{C}_a$ is an anti-clockwise contour enclosing the
origin with $|w| > |\Theta| \sqrt{\tilde{q}_r/\tilde{q}_l}$ and
$\mathcal{C}_b$  goes clockwise around the origin with $|w| <
\sqrt{\tilde{q}_r/\tilde{q}_l}/|\Theta|$. We now expand the exponentials in a
Laurent series and integrate each term to finally arrive at Eq.~(\ref{final})
upon resumming the resulting series.

The main issue raised by our results is that of the origin of the
distinction between the semiclassical and form-factor results.
Both methods make subtle unproven assumptions about orders of
limits: the commuting of the $t \rightarrow \infty$ limit with
either the semiclassical (in which the low momentum limit is taken
in obtaining Eq.~(\ref{smatrix})) or the form factor expansions.
It could be that such an assumption is invalid in one of the
methods, and only the other result is generically valid in the
long-time limit. However, we suspect that the distinction reflects
a fundamental physical difference between generic and integrable
systems, and the two results apply in their respective cases.

After this work was completed, we became aware of independent work
by Rapp and Zarand\cite{Zarand} who find that diffusive behavior
appears in the correlators of the one-dimensional $Q$-state
quantum potts model for $Q>2$. Remarkably, the correlation
functions in the two cases differ only in the choice of the
parameter $\Theta$ which takes the
value $\Theta = -1/(Q-1)$ in the Potts case.

We would like to acknowledge A.~Tsvelik for useful discussions
regarding the results obtained in
Ref~\onlinecite{AltshulerTsvelik}, and thank G.~Zarand for useful
discussions regarding the the results of Ref~\onlinecite{Zarand}.
One of us (KD) would like to thank A.~Dighe and R.~Raghunathan for
useful discussions, and D.~Dhar for insightful discussions and
suggesting the trick used in arriving at an integral
representation of Eq.~(\ref{e8}). S.S. was supported by NSF Grant
DMR-0537077.


\end{document}